\documentclass{article}
\usepackage{spconf,amsmath,graphicx}
\usepackage{hhline}
\usepackage{amsfonts}
\usepackage[table]{xcolor}

\title{MAST: Multiscale Audio Spectrogram Transformers}
\name{Sreyan Ghosh$^{1\star}$, Ashish Seth$^{2\star}$, S. Umesh$^{2}$, Dinesh Manocha$^{1}$
\thanks{\hspace*{-1mm}$^{\star}$These authors contributed equally to this work}}
\address{
    $^1$University of Maryland, College Park, USA\\
  $^2$Speech Lab, Department of Electrical Engineering, IIT Madras, Chennai, India\\
  }
\begin{document}
%
\maketitle
\begin{abstract}
We present Multiscale Audio Spectrogram Transformer (MAST) for audio classification, which brings the concept of multiscale feature hierarchies to the Audio Spectrogram Transformer (AST) \cite{gong2021ast}. Given an input audio spectrogram, we first \emph{patchify} and project it into an initial temporal resolution and embedding dimension, post which the multiple stages in MAST progressively expand the embedding dimension while reducing the temporal resolution of the input. We use a pyramid structure that allows early layers of MAST operating at a high temporal resolution but low embedding space to model simple low-level acoustic information and deeper temporally coarse layers to model high-level acoustic information with high-dimensional embeddings. We also extend our approach to present a new Self-Supervised Learning (SSL) method called SS-MAST, which calculates a symmetric contrastive loss between latent representations from a student and a teacher encoder, leveraging \emph{patch-drop}, a novel audio augmentation approach that we introduce. In practice, MAST significantly outperforms AST by an average accuracy of 3.4\% across 8 speech and non-speech tasks from the LAPE Benchmark \cite{9868132}, achieving state-of-the-art results on keyword spotting in Speech Commands. Additionally, our proposed SS-MAST achieves an absolute average improvement of 2.6\% over the previously proposed SSAST \cite{gong2022ssast}\footnote{https://github.com/Sreyan88/LAPE}.
\end{abstract}
\section{Introduction}
\label{sec:introduction}


Natural signals such as speech and audio are hierarchically structured across various different timescales, spanning tens (e.g., phonemes) to hundreds (e.g., words) of milliseconds. Each type of signal is highly variable and context-dependent \cite{de2017hierarchical}. It has been shown that deep neural networks (DNNs)  can excel at learning these complex temporal structures from natural signals. Explicitly modeling multiscale features in the time domain for processing speech has been extensively studied for tasks like Speech Emotion Recognition (SER) ~\cite{sivanagaraja2017end}, Speaker Verification (SV) ~\cite{zhu2020vector}, TTS~\cite{li2021towards} and ASR ~\cite{zhu2016learning}, and has produced favorable results. Another excellent example is the five-layer convolutional wav2vec 2.0 \cite{baevski2020wav2vec} feature extractor relatively small strides (\{5,4,2,2,2\}), which gradually decreases the embedding space in the time dimension.

\begin{figure}[t]
  \centering
  \includegraphics[width=0.48\textwidth]{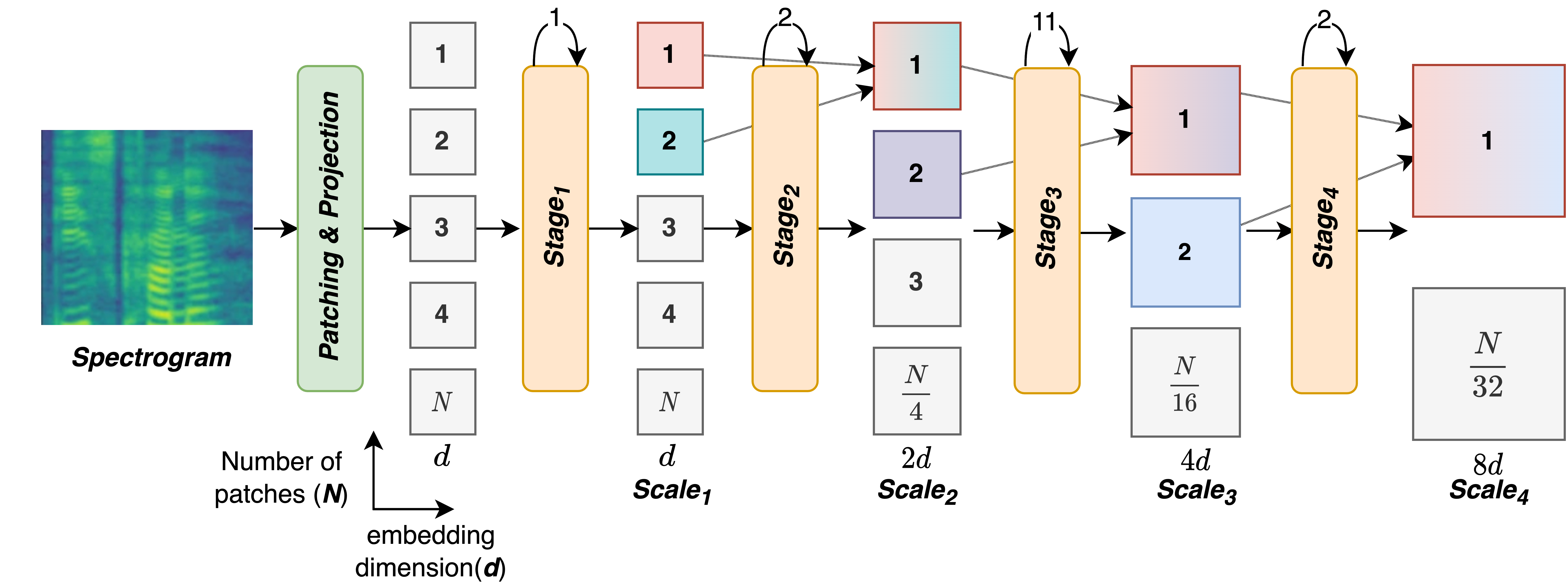}
  \caption{\small We show how a \emph{patchified} audio spectrogram is processed across multiple stages in MAST. Each stage in MAST is made up of multiple transformer blocks (1,2,11,2). Every stage progressively increases the patch embedding dimension by $2\times$ while decreasing the number of patches by $1/4^{th}$. Different colors in each scale represent how the convolution pooling operator convolves patches across multiple stages.}
  \label{fig:figure_0}
\end{figure}

\begin{figure*}[t]
  \centering
  \includegraphics[width=1.0\textwidth]{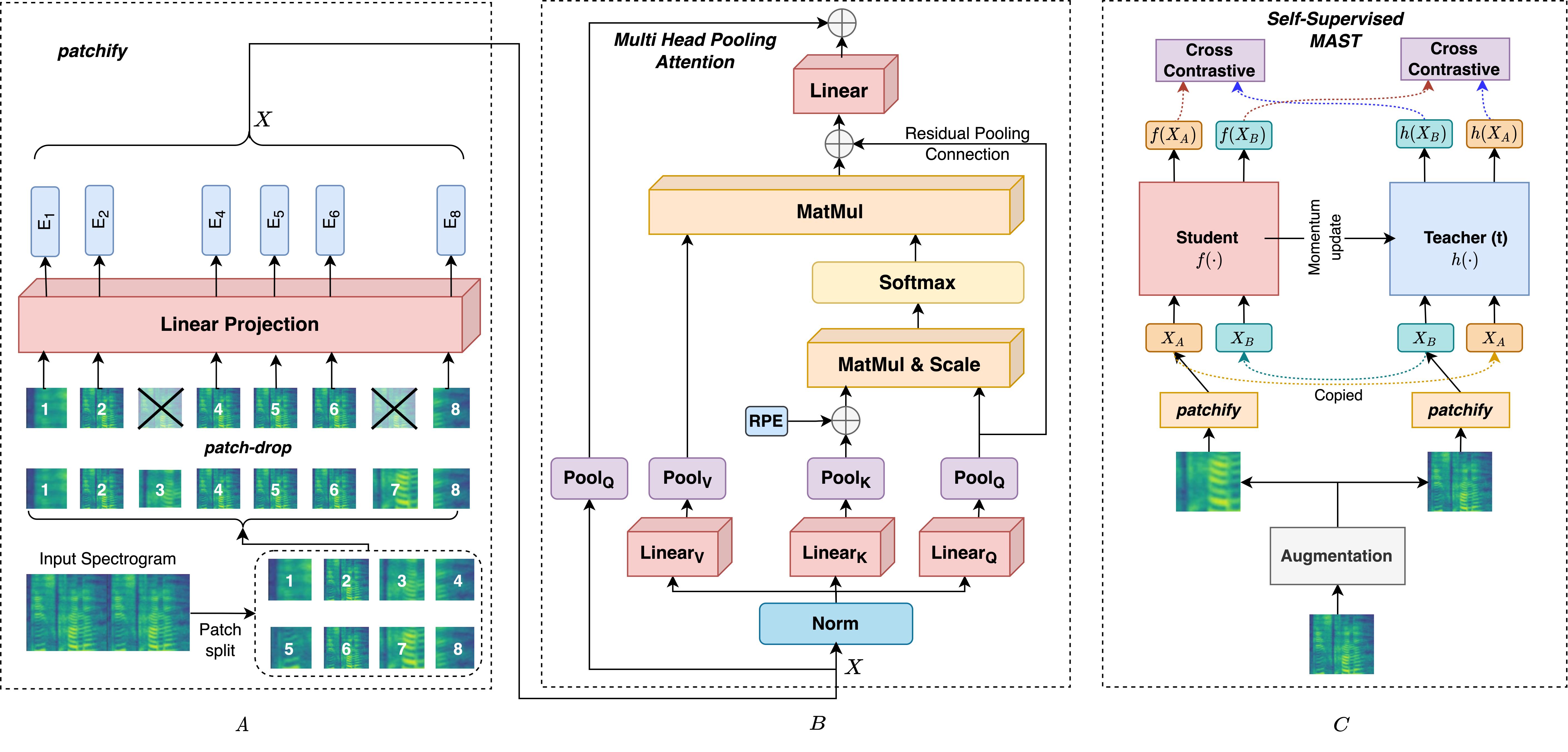}
  \caption{\small Illustration of the Multiscale Audio Spectrogram Transformer (MAST): (A) The input audio is first transformed to log-scaled mel-spectrogram before it is \emph{patchified}. Optionally, we randomly drop $j\%$ for the patches only for SSL pre-training. (B) The \emph{patchified} log-mel-spectrogram is passed through multiple stages of MAST each having multiple transformer blocks with Multi Head Pooling Attention at its core. (C) SS-MAST: For SSL pre-training of MAST we make 2 copies of randomly augmented log-mel-spectrogram and solve a cross-contrastive loss between the student and the momentum-teacher networks.}
  \label{fig:figure_1}
\end{figure*}

To confirm our hypothesis on hierarchically structured natural signals, we highlight a key architectural design choice common across the best performing CNN-based architectures for audio classification in literature. With a spectrogram as input, these models keep decreasing the temporal resolution while increasing the frequency embedding dimension progressively as the input is processed from lower to higher layers in the network \cite{niizumi2021byol,wang2022towards,kazakos2021slow}. This design choice for \emph{pure-CNN} models allows them to hierarchically learn simple low-level acoustic features in the lower stages aided by high temporal and low embedding dimensions to complex high-level acoustic features in the higher stages aided by low temporal and high embedding dimensions. Though CNN models have dominated the audio classification space for the last decade, the recently introduced AST \cite{gong2021ast}, which was the first pure-attention \emph{non-CNN} audio classification model, outperformed \emph{pure-CNN} models in a variety of audio classification tasks. AST is built on Vision Transformers (ViT) \cite{dosovitskiy2020image} and shares the exact same configuration with ViT. Similar to ViTs, where the input image is divided into several patches (also known as \emph{patchifying}), AST first \emph{patchifies} the raw audio spectrogram before passing it through the various stages in AST, each of which consists of multiple transformer blocks. However, one drawback of AST is that it fails to leverage CNN-like feature hierarchies, which has benefited \emph{pure-CNN} models for audio classification in the past decade.
\vspace{1mm}


{\noindent \bf Main Contributions:} (1) We introduce MAST, a new \emph{convolution-free} transformer-based architecture for audio classification. MAST builds on AST \cite{gong2021ast} and modifies the AST architecture to incorporate the idea of multiscale feature hierarchies into it. MAST differs from AST primarily in the design of the transformer blocks, wherein the early stages of MAST operate at a high temporal resolution to focus on modeling simple low-level properties in the audio signal, including acoustic, prosodic, and spectral cues aided by low embedding dimension (see Fig. \ref{fig:figure_0}). This helps MAST to focus on local temporal patterns like speed, energy, pitch, pause, etc., in the transformer layers in the early stages. Moreover, the transformer layers in the deeper stages can effectively focus on temporally coarse but complex high-level acoustic properties like semantics, emotion, etc. This core concept of multi-scale stages in MAST is employed by using the Multi Head Pooling Attention (MHPA) layers \cite{fan2021multiscale,li2022mvitv2} in place of Multi Head Attention (MHA) layers in AST. Further, MAST also reduces the quadratic self-attention complexity in AST~\cite{gong2021ast} due to lower temporal resolution in higher stages. Finally, MAST outperforms AST in both speech and non-speech tasks from the LAPE Benchmark \cite{9868132}.
(2) In addition to MAST, in this paper, we also present SS-MAST, a new SSL approach that helps MAST achieve higher performance in low-resource supervised learning settings by learning general-purpose audio representations from unlabeled audio. SS-MAST differentiates itself from other SSL-based audio representation learning approaches in literature by virtue of its design by which it overcomes the instability issue that has been seen to impact self-supervised ViT training \cite{chen2021empirical}, does not solve a Masked Acoustic Modeling (MAM) task like SSAST \cite{gong2022ssast} and is also more flexible in terms of usage across different architectures (see Section \ref{sec:related} for explanation). SS-MAST solves a symmetric instance-level cross-contrastive task between student and teacher encoders, wherein each encoder is fed with two randomly augmented versions of the same audio, and the model tries to and for each positive instance sampled from an encoder output batch, negative instances are sampled from the output of a different encoder for a different augmentation. Our proposed SS-MAST performs better than SSAST on both AST and MAST in the LAPE pre-training and fine-tuning setup.

\section{Related Work}
\label{sec:related}
In the past, the use of SSL to learn general-purpose audio representations has been achieved by solving either contrastive learning \cite{wang2022towards,9868132}, clustering \cite{ghosh2021deep}, or reconstruction \cite{niizumi2021byol}. Unlike MAM in speech, SSL in audio generally tries to make representations learnt augmentation invariant, with recent exceptions being [18, 19]. One common trait among most of these approaches is all of them that they all use CNNs, with the only exception being SSAST \cite{gong2022ssast}, which employs a joint discriminative and generative masked spectrogram patch modeling task on the AST. We acknowledge the fact, with strong evidence from literature \cite{new_berkley}, that MAM tasks are better suited for tasks like ASR and phoneme recognition by the nature of their design. Adding to this, SSAST requires equal temporal resolution in input and output spaces which restricts it's use to only AST backbone encoder in audio. MAST is inspired from MViT \cite{fan2021multiscale,li2022mvitv2}, first developed for image and video recognition, and remarkable performance in video recognition with strong implicit temporal bias. For the same reason, MAST benefits over AST for audio classification tasks due to the presence of natural hierarchical time-domain signals in audio.

\section{Multiscale AST}
\label{sec:mast}


In the next sub-sections we describe the individual components in MAST which differentiates it from AST, including MHPA transformer blocks, a residual query connection, and relative position embedding (RPE). We illustrate these components in Fig. \ref{fig:figure_1} (Part B).

\subsection{Multi Head Pooling Attention}
\label{sec:arch}
To achieve varying spatiotemporal resolution through progressive stages in MAST, we use MHPA-equipped transformer blocks. Unlike vanilla MHA, where the embedding dimension and the temporal resolution remain fixed, MHPA pools the sequence of latent tensors to reduce the sequence length along the time axis. Formally put, let $X$ be an input to the block with sequence length $L$. Like MHA, MHPA first projects $X$ to key, query, and value tensors $\hat{K}, \hat{Q}, \hat{V} \in \mathbb{R}^{L \times D}$. Next, an intermediate pooling step pools these tensors in sequence length to obtain $K, Q, V \in \mathbb{R}^{\Tilde{L} \times D}$ where $\tilde{L}=\left\lfloor\frac{L+2 p-k}{s}\right\rfloor+1$ and $p$, $s$, and $k$ are the padding, stride, and kernel sizes of the convolution pooling operator, respectively. Though the model is flexible enough to use any kind of intermediate pooling operator \textbf{Pool}, we empirically find that a convolution operator with overlapping kernels works best, which is re-confirmed by strong evidence of the use of convolution subsampling layers in the ASR literature. In practice, though the output sequence length of each block and each transformer layer output can be configured flexibly by changes to stride $s$ in the pooling operator \textbf{Pool}, for $\hat{Q}$ we pool with $s > 1$ at only the first transformer block of each stage and constrain $s = (1,1,1)$ across the transformer layers of the same block. For $\hat{K}$ and $\hat{V}$, since both their sequence lengths be the same, we constrain $s$ to be the same for both $\hat{K}$ and $\hat{V}$ throughout a block but vary $s$ adaptively across stages. This setting is similar to the original MViT \cite{fan2021multiscale}. More details about the output of each stage can be found in Fig. \ref{fig:figure_0}. Finally, after calculating self-attention \cite{vaswani2017attention} over the shortened vectors $\hat{K}, \hat{Q}$, and $\hat{V}$, we also apply a residual pooling connection with the pooled Q tensor to increase information flow and facilitate training. The multi-layer perceptron (MLP) after the MHPA in the transformer block at the end of each stage is responsible for increasing the output patch embedding dimension.

\subsection{Relative Position Embedding}
\label{sec:arch}
For MAST, we drop the absolute positional embedding for patches in the input stage and adopt the relative position embedding proposed in \cite{li2022mvitv2}. This overcomes the problem wherein two patches change with their absolute position in mel-spectrograms even if their relative positions stay unchanged. This becomes extremely important in speech tasks like SER, wherein high-level features like semantics depend a lot on the relative position of excitation in mel-spectrograms. We encode the relative position between the two input elements, $i$ and $j$, into positional embedding $R_{p(i)} , R_{p(j)} \in \mathbb{R}^{D}$ in the MHPA blocks:

\vspace{-0.5cm}

\begin{equation}
    \operatorname{Attn}(Q, K, V)=\operatorname{Softmax}\left(\left(Q K^{\top}+E^{(\text {rel })}\right) / \sqrt{d}\right) V
\end{equation}
where $E_{i j}^{(\mathrm{rel})}=Q_i \cdot R_{p(i), p(j)}$. Thus, by design, MAST reduces the time complexity in MHA, which scales quadratically with sequence length.

\begin{table*}[t]
\centering
\small
\caption{\small Result comparison of MAST and SS-MAST with AST and SSAST on 8 different speech and non-speech tasks. Best scores for each sub-section are shown in bold. \emph{p.d.} indicates \emph{patch-drop}. As we see, MAST outperforms AST in all settings.}
\vspace{1mm}
\begin{tabular}{|l l| c c c c c c c c c|}
\hline 
\textbf{Model} & \textbf{Initialization} & \textbf{SC-V1} & \textbf{SC-V2(12)} & \textbf{SC-V2(35)} & \textbf{VC} & \textbf{VF} & \textbf{IC} & \textbf{NS} & \textbf{US8K} & \textbf{Avg.}\\
\hhline{|= =| = = = = = = = = =|}

AST & \emph{random} & 87.3 & 88.2 & 92.7 & 30.1 & 72.3 & 51.9 & 70.9 & 50.1 & 67.9\\

AST & \emph{ImageNet weights}& 90.0 & 91.1 & 93.1 & 51.2 & 79.8 & 54.2 & 71.1 & 62.3 & 74.1 \\

AST & \emph{ImageNet + SSAST} & 95.5 & 94.2 & 94.4 & 53.3 & 84.4 & 58.8 & 74.3 & 78.0 & 79.1\\

AST & \emph{SS-MAST}& 92.0 & 91.8 & 94.0 & 44.9 & 81.8 & 55.2 & 72.1 & 68.3 & 75.0\\

AST & \emph{ImageNet + SS-MAST} & 96.0 & 94.4 & 95.4 & 53.4 & 88.8 & 60.1 & 76.4 &79.3 & 80.5\\

\hhline{|= =| = = = = = = = = =|}

MAST &\emph{random}& 91.0 & 92.2 & 93.4 & 33.2 & 74.3 & 58.3 & 73.4 & 54.4& 71.3\\

MAST & \emph{ImageNet weights} & 92.0 & 93.1 & 94.2 & 54.4 & 87.3 & 61.0 & 75.4 & 64.4 & 77.7\\

MAST & \emph{SS-MAST} & 93.2 & 94.5 & 95.0 & 54.9 & 88.3 & 62.2 & 76.8 & 74.2 & 79.8\\

MAST & \emph{ImageNet + SS-MAST} &  97.0 & 96.8 & 96.4 & 56.7 & 89.2 & 64.0 & 80.6 & 84.0 & 83.1\\

MAST & \emph{ImageNet + SS-MAST + p.d. } & \cellcolor{blue!9} \textbf{97.4} & \cellcolor{blue!9} \textbf{96.8} & \cellcolor{blue!9} \textbf{96.6} & \cellcolor{blue!9} \textbf{57.3} & \cellcolor{blue!9} \textbf{90.0} & \cellcolor{blue!9} \textbf{64.4} & \cellcolor{blue!9} \textbf{81.2} & \cellcolor{blue!9} \textbf{84.8} & \cellcolor{blue!9} \textbf{83.6}\\

\hline

\end{tabular}
\label{table:results}
\end{table*}

\section{Self-Supervised MAST}
\label{sec:ss-mast}

\subsection{Algorithm}
\label{subsec:ss-mast_algo}

We employ contrastive learning between output representations from the student and the momentum teacher. This paradigm was first introduced by MoCo \cite{he2020momentum} and adapted for audio by \cite{9868132}. In brief, both the student $f(.)$ and the teacher encoders $h(.)$ are fed with batches of different augmentations $x^a_i$ and $x^b_i$ respectively of the same audio sample $x_i \in X$, where $X$ is a batch of size $N$ indexed by $\{0, \cdots, i, \cdots ,N-1\}$. For each audio representation $f(x^a_i)$ from the student, the corresponding teacher representation of the same but differently augmented audio sample $h(x^b_i)$ acts as a positive and teacher representations $\tilde{x_{i}}$ of other audio samples act as negatives. Thus, the contrastive loss can be defined as follows:

\vspace{-0.5cm}
\setlength{\abovedisplayskip}{3pt}

\begin{multline}
\label{eqn:info}
\mathcal{L}_{\text {InfoNCE}}(f,h)=-\log(\\
\frac{\exp \left(f\left(x^{a}_{i}\right) \cdot h\left(x^{b}_{i}\right) / \tau\right)}{\exp \left(f\left(x^{a}_{i}\right) \cdot h\left(x^{b}_{i}\right) / \tau\right)+\sum_{i=0}^{K} \exp \left(f\left(x^{a}_{i}\right) \cdot h\left(\tilde{x_{i}}\right) / \tau\right)})
\end{multline}
\setlength{\belowdisplayskip}{3pt}

However, we differ from \cite{9868132} in 2 major ways: 1) We do not use a queue in our setup, which we empirically find to be not improving results when accompanied by strong augmentations. 2) We use a symmetrized version of the contrastive loss, where we calculate a cross-contrastive loss between the student and teacher representations, and our final loss $\mathcal{L}_{\text {InfoNCE}}$ is computed by $\mathcal{L}_{\text {InfoNCE}}$ = $\mathcal{L}_{\text {InfoNCE}}(f,h)$ + $\mathcal{L}_{\text {InfoNCE}}(h,f)$.

\subsection{Augmentations}
\label{subsec:ss-mast_aug}

SSL algorithms, which solve a task such that augmentations are invariant, depend heavily on the augmentation quality (MAM and MLM being exceptions). For our setup, we use 3 different types of augmentation schemes, namely 1) RRC, 2) mixup and 3) \emph{patch-drop} (proposed). For more details about RRC and mixup, we refer our readers to \cite{niizumi2021byol}. In this paper, we propose patch-drop, which randomly drops $20\%$ of patches from the patched log-mel-spectrogram input before it is input to MAST. In Table \ref{table:results}, we empirically show that patch-drop improves SSL-based audio representation learning when used together with RRC and mixup.

\section{Implementation Details}
\label{sec:implementation}

{\noindent \textbf{Dataset.}} The LAPE Benchmark \cite{9868132} is the first proposed benchmark for evaluating learned audio representations across a mix of different speech and non-speech tasks. In all our experiments, we follow the exact SSL upstream setting from LAPE and evaluate our learned representations on task-specific downstream fine-tuning settings with tasks from LAPE. To be precise, we do SSL pre-training on a combination of the 10\% of AudioSet and the entire FSD50K and evaluate learned representations on VoxCeleb (VC) \cite{Nagrani_2017} for speaker identification, Speech Commands (SC) v1 and v2 \cite{warden2018speech} for keyword spotting, VoxForge (VF) \cite{Voxforge.org} for language identification,  IEMOCAP (IC) \cite{busso2008iemocap} for speech emotion recognition, NSynth \cite{engel2017neural} for musical note instrument classification, and finally US8K \cite{10.1145/2647868.2655045} for acoustic event classification.
\vspace{1mm}

{\noindent \textbf{Experimental Setup.}} For all our experiments, we follow the MViT-\emph{Base} architecture for MAST and ViT-\emph{Base\textsubscript{384}} architecture for AST. We find the smaller variant outperforming AST and would like to explore bigger models in future work. For fine-tuning all variants of AST and MAST, we find a learning rate of $3e^{-4}$ and a batch size of 64 to be optimal. For pre-training SS-MAST, we add  an extra linear layer on top of each encoder to project the output to $\mathbb{R}^{256}$ and train it with a batch size of 512 and a learning rate of $3e^{-4}$. We find a \emph{patch-drop} frequency of $j=20\%$ to work best for us among \{10, 20, 30, 40, 50\}.
\vspace{1mm}


\section{Results and Analysis}
\label{sec:results}

Table \ref{table:results} compares the performance of both AST and MAST under various settings. We show results for all 8 tasks on randomly initialized weights without any SSL pre-training(\emph{random}), initialized with ImageNet fine-tuned weights (\textit{ImageNet weights}), and with further SS-MAST pre-training (\textit{ImageNet + SS-MAST}). MAST significantly outperforms AST on all these 3 settings with an average gain of 3.2\%. Additionally, to prove the superiority of SS-MAST in isolation, we pre-train AST with SS-MAST pre-training on AudioSet without ImageNet weights (\textit{SS-MAST}) and also compare both SSL pre-training algorithms with ImageNet weights (\textit{ImageNet + SS-MAST vs ImageNet + SSAST}). SS-MAST pre-training achieves an average 1.4\% improvement over SSAST pre-training. We do not implement SSAST on MAST as MAST does not maintain equal dimensions in the input and output embedding space. Our best-performing system is SS-MAST pre-trained on LAPE with our proposed \emph{patch-drop} augmentation, which boosts performance by an average of 0.5\%.

\section{Conclusion and Future Work}
\label{sec:conclusion}

In this paper, we present Multiscale AST. Due to its strong implicit temporal bias, we hypothesize MAST is better suited for audio and speech classification tasks. We empirically show that MAST outperforms AST for various speech and non-speech tasks. We also present SS-MAST, a new SSL-based audio pre-training methodology for MAST. SS-MAST outperforms SSAST when employed with either AST or MAST. 



\newpage
\bibliographystyle{IEEEbib}
\bibliography{strings,refs}

\end{document}